\documentclass[prb,preprint,longbibliography]{revtex4-2} 

\usepackage{amsmath}  
\usepackage{amsfonts}
\usepackage{breqn}
\makeatletter
\begingroup\makeatletter
\catcode`,=\active
\global\let\breqn@comma,
\protected\gdef,{\ifmmode\expandafter\breqn@comma\else\expandafter\active@comma\fi}
\endgroup
\makeatother
\usepackage{graphicx}
\usepackage{epstopdf}
\usepackage{xcolor}
\usepackage{url}
\begin{document}

\title{A new (old?) coupled pendulum experiment.}

\author{Neil G.~R.~Broderick}
\email{n.broderick@auckland.ac.nz} 
\author{Benjamin Pollard}
\email{ben.pollard@auckland.ac.nz}
\author{Karthik Sivasubramanian}
\author{Marc Lescano}
\affiliation{Department of Physics, University of Auckland, Auckland NZ}
\author{Chengjie Chen}
\email{chenchengjie77@163.com}
\affiliation{School of Electrical and Information Engineering, Jiangsu University of Technology, Changzhou 213001, China}

\date{\today}

\begin{abstract}
We present a description and analysis of a coupled pendulum experiment that has been 
used at the University of Auckland for many years. Although similar experiments are very common in undergraduate physics labs, both the physical system and our teaching approach are, we believe, novel. Compared to other coupled pendulum experiments, it has the advantage that the coupling can easily be varied, allowing students to observe the changing nature of the normal modes and beat frequencies. Additionally, the system displays a surprising degree of complexity, making it of interest to more advanced students in laboratory courses that are augmented with computer-based numerical modeling and algebra software. 
\end{abstract}

\maketitle 

\section{Introduction}

Sidney Coleman once remarked that
``The career of a young theoretical physicist consists of treating the harmonic oscillator in ever-increasing levels of abstraction.'' \cite{coleman} 
While this 
quotation 
referred to a single harmonic oscillator, it could just as well apply to coupled oscillators, which form the basis of much of many-body physics from crystals 
to the mechanics of walking.\cite{stewart1} Students are often introduced to coupled oscillators in an undergraduate experiment in which they study the beats 
and normal modes in a system such as coupled pendula. These experiments are normally sufficiently complex as to be appropriate for students in 
beyond-first-year laboratory sessions, where learning objectives should include 
more sophisticated aspects of interpreting data\cite{AAPTLabs}, experimental modeling\cite{Dounas-Frazer2018a}, and computational 
thinking\cite{Behringer2017, AAPTComputation}, all of which are increasingly important skills for preparing students for post-graduate career paths and 
advanced-degree research\cite{Heron2016, McNeil2017}.

A typical coupled pendulum consists of two identical pendula coupled 
by a spring. Students work with the 
linearised equations of motion:
\begin{eqnarray}
	\ddot x_1 & = -\omega_0^2\, x_1 -\frac{\kappa}{m} (x_1-x_2) \\
	\ddot x_2 & = -\omega_0^2\, x_2 - \frac{\kappa}{m} (x_2-x_1),  \label{scp1}
\end{eqnarray}
where $x_1$ and $x_2$ describe the positions of the pendula's bobs (both of mass $m$ and length $l$), $\omega_0 =\sqrt{g/l}$ is the uncoupled 
frequency, and $\kappa$ is the spring constant.
The aim of the experiment is to find the frequencies of the normal modes by (a) measuring them directly and (b) measuring the beat frequencies, and 
then show that the two methods agree. For the system described by these expressions, the normal modes correspond to the symmetric (even) mode
 [$x_1(t)=x_2(t)$] which has frequency $\omega_0$, and the anti-symmetric (odd) mode  [$x_1(t)=-x_2(t)$] which has frequency  $\sqrt{\omega_0^2+2 \kappa/m}$.

A drawback of most of the setups used in these experiments is the difficulty of changing the coupling constant $\kappa$. This in turn makes it hard to verify the 
prediction for the frequencies of the normal modes and in particular whether the frequency of the anti-symmetric mode follows the predicted square root relationship. 
Here we describe  a coupled pendulum activity designed for an advanced physics laboratory that overcomes this limitation. The experiment employs a novel  
setup that is inexpensive 
and easy to construct, making it accessible for a wide range of educational settings. Also, by incorporating  use of computer algebra software to aid in the analysis, we aim 
to emphasize the increasing importance of cultivating ``computational thinking'' in the physics 
curriculum\cite{Behringer2017, AAPTComputation, Caballero2015, Lane2021, PICUP, Phillips2023}.
In particular, students acquire experience with  packages such as Maple and Mathematica to help them  derive and solve equations that are too complex or 
time-consuming to be analysed by hand. Beyond-first-year physics lab courses can provide a natural context for linking  computational thinking with experimental skills.

In this paper we first describe and then analyse the coupled pendulum experiment used in the Advanced Lab at the University of Auckland. The analysis starts with 
the simple description currently used and then we present a more accurate analysis using a Lagrangian framework. From the equations of motion, we can find the
frequencies of the normal modes, allowing accurate predictions to be made and compared to experiments. We also present possible extensions to the experiment
suitable for either theoretically or practically -- minded students.

\section{Coupled Pendulums in Our Curriculum}
Within our department, the role of experimental teaching and that of our ``advanced labs'' is twofold. First, we aim to develop lab-specific skills needed for students' future careers, including those who wish to undertake a graduate degree. Second, we try to align with the third-year physics courses by providing experiments that are directly related to the subjects being taught. Over the course of the semester, students typically undertake four or five experiments, with each taking between 8 - 10 hours to complete, including 
preparation and the analysis of the results. 
They are assessed via both  an oral discussion and by submitting a ``deliverable,''  typically a written report. There are typically 10 - 20 students each semester. We have a broad definition of ``experiment,'' with one or two  being $100\%$ computer-based 
while others focus on  students developing particular experimental skills. Most include 
numerical modeling to compare  experimental results with predictions from established theory. 
The coupled pendulum experiment is designed to draw on students' knowledge of advanced mechanics from a third year course that focuses on Langrangian and nonlinear dynamics. Students learn to use Tracker software\footnote{https://opensourcephysics.github.io/tracker-website/} to obtain precise frequency measurements and also gain computational skills in symbolically manipulating  Langrangians to obtain experimental predictions.

\begin{figure}[h]
\vspace{0.5cm}
\centering
\includegraphics[width=4.5cm]{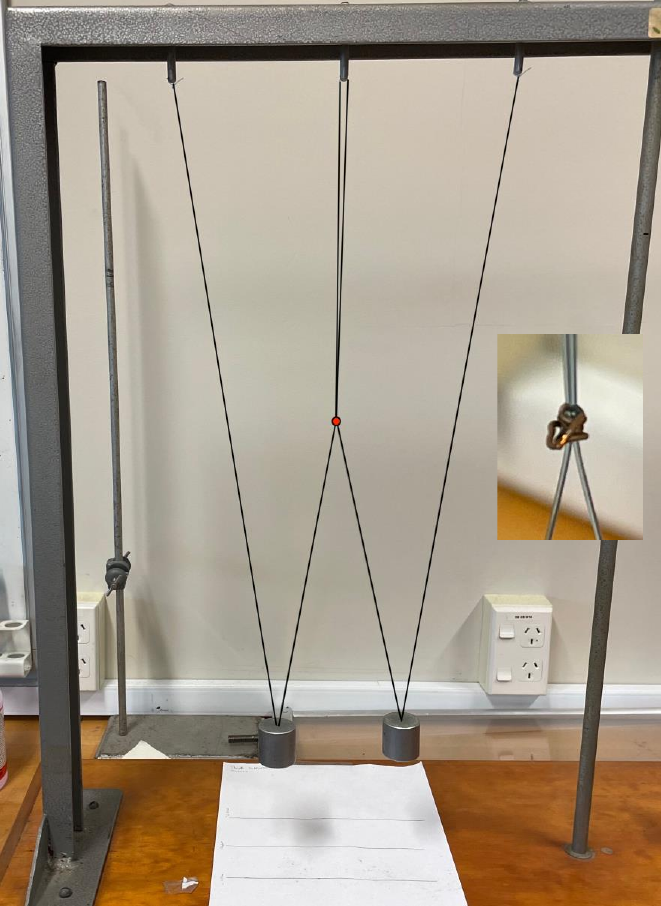}
\includegraphics[width=10cm]{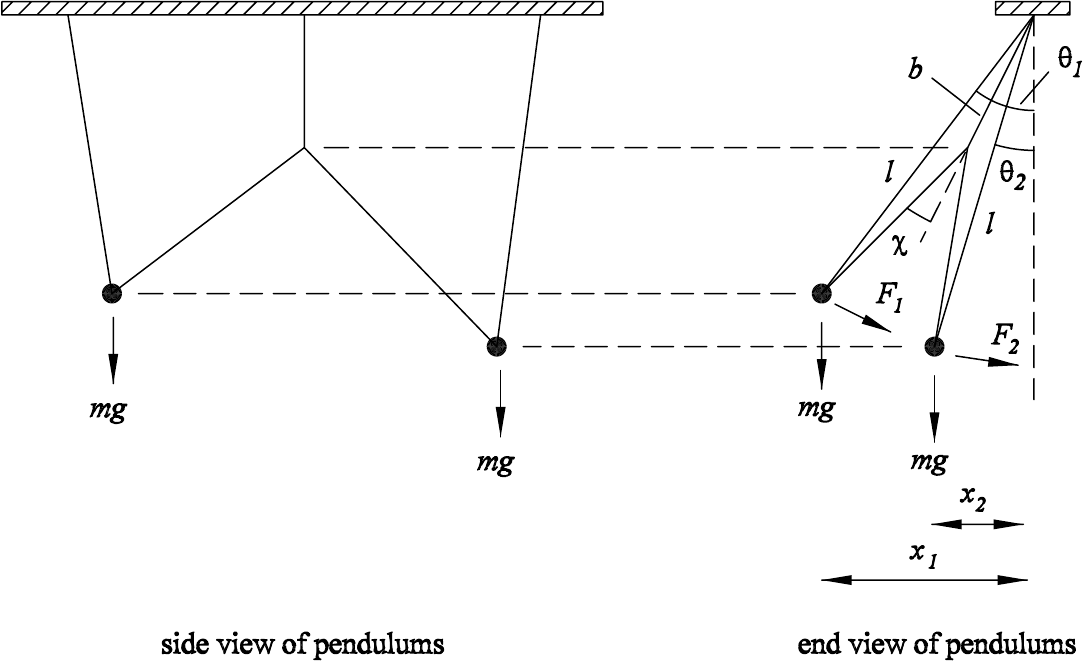} \\ \vspace{0.2cm}
\includegraphics[width=15cm]{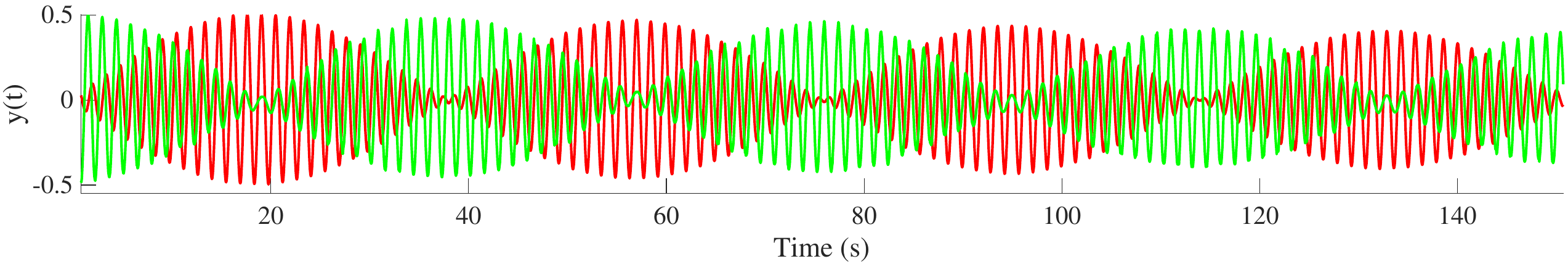}
\caption{(Color online)  (a) Picture of the coupled pendulum experiment at the University of Auckland. We have superimposed black lines over the fishing line used 
in order to improve the contrast.  Similarly, we indicate the position of the twisted wire via a red dot to enhance visibility, with the insert showing a magnified
view of the twisted wire. (b) Schematic of the experiment taken 
directly from the current  lab manual (2024). The central picture shows the situation when the bobs are in motion, which is why the lengths of the strings appear
different. The bottom picture shows a time trace of the positions of the pendulum bobs  recorded with a video camera, clearly 
showing the beating of the modes. The right 
pendulum bob is shown in red and the left bob is in green.}
\label{auckland}
\end{figure}

Figure  ~\ref{auckland} shows the setup of our experiment.  On the left is a picture of the system and on the right is a schematic illustration taken 
from the current lab manual. 
Both pendulum bobs are attached to a common rail by two strings of length $l$. One support point for each  bob is fixed to the centre of the apparatus 
(taken to be the origin in 
the later analysis), while the second  is symmetrically positioned at a distance  $d$ from the centre. At rest and uncoupled, the tops of the bobs are at a distance 
$l_{\textrm{eff}}=\sqrt{l^2-(d/2)^2}$ below the rail and thus for small oscillations have periods $w_0 \approx\sqrt{g/l_{\textrm{eff}}}$. This expression 
is exact for point masses, but for extended masses such as the ones used by as a correction is needed (which is a subject for future analysis). 
The coupling is provided by fixing the two centre support strings together at a distance $b$ below the rail. This is done by using a short twisted piece of wire 
that can be slid up or 
down to quickly alter the coupling strength. When $b=0$, the pendula are uncoupled; increasing $b$ increases the coupling strength. 
In order to minimise any unwanted coupling through the frame, the apparatus was constructed from steel with typical cross-sectional dimensions of 4 - 5 mm; the age of the
apparatus is such that we do not have the original dimensions. 

During the experiment, students vary the coupling length $b$ then displace one or the other of the weights out of the plane of the picture and observe 
the resulting oscillations. 
Along the bottom of Fig.~\ref{auckland} 
is a  time trace of the motion of the  bobs recorded with a video camera and analysed using Tracker software.\cite{Brown.2009} This trace shows the presence of the 
beating between the normal modes of the system.

The derivation of the relevant equations presented in the current lab handout is very brief and consists of the following: \\
``For the two coupled pendulums shown in Fig. \ref{auckland}, assuming the angles are small and the tensions in the strings are equal, the force $F_{1}$ is given by
\begin{equation}
F_{1}=-\frac{mg}{2}\theta_{1}-\frac{mg}{2}\left(  \frac{\theta_{1}+\theta_{2}%
}{2}+\chi\right).
\end{equation}

Using $\theta_{1}\approx x_{1}/l$, $\theta_{2}\approx x_{2}/l$ and 
$\chi\approx\left(  x_{1}-x_{2}\right)  /\left(2\left(  l-b\right)\right)  ,$ this can be conveniently written as
\begin{equation}
F_{1}\approx-\frac{mg}{l}x_{1}-\frac{mgb}{4l\left(  l-b\right)  }\left(
x_{1}-x_{2}\right).
\end{equation}

The expression for $F_{2}$ is similar. Notice that the forces are still linearly related to the positions, so the two coupled pendula are another example of a linear system. 
The resulting equation of motion is
\begin{equation}
\label{simple}
\frac{\mathrm{d}^{2}}{\mathrm{d}t^{2}}\left(
\begin{array}
[c]{c}%
x_{1}\\
x_{2}%
\end{array}
\right)  =\left(
\begin{array}
[c]{cc}%
-\omega_{0}^{2}-k & k\\
k & -\omega_{0}^{2}-k
\end{array}
\right)  \left(
\begin{array}
[c]{c}%
x_{1}\\
x_{2}%
\end{array}
\right),
\end{equation}
where $\omega_{0}^{2}$ is $g/l$ and $k$, a constant, is given by $k={gb}/{4l\left(  l-b\right)  }$.''
These equations empirically model the data reasonably well, but we have not been able to justify their derivation, so we decided to derive an equation that more fully incorporates the complexity of the apparatus.

Despite the simplicity of this arrangement and the ease with which the coupling can be changed, 
this system appears to be unique; we would be very interested in knowing if any reader has ever seen a similar setup. We know that it has existed in its 
current form since the 1970s and was old even then.  While both our experiment and others have as the main objective the measurement of the beat 
frequency, the ease with which the coupling constant can be changed means that a more thorough investigation is possible within a limited time. Sliding the 
metal clip up and down can even be done while the pendula are in motion, allowing students to see the dynamical effect of changing the coupling constant, 
a procedure  not possible using other systems. As we show in section \ref{extensions}, 
the system can be extended to multiple coupled pendulums, also something we have not seen done elsewhere.


\section{Improved theoretical analysis of the coupled pendulum}
\subsection{Derivation of the Lagrangian and generalised co-ordinates}

The analysis of the coupled pendulum presented at the start of the lab manual is very brief and reflects the time when the experiment was designed 
and built (likely late 1960s).  For students doing the experiment without the aid of digital measuring devices and computers, this analysis is sufficient 
in that it explains the underlying physics and produces a theoretical model that agrees with experimental measurements. However, the derivation is 
ambiguous in that it uses $l$ for both the length of string joining the bobs to the support and for the length of the pendulum itself; that is, it neglects the 
$2d$  offset support of the bobs.  While the error that comes from neglecting the offset distance is small in the current setup,
 in some realisations $d$ could be quite large, rendering the previous analysis incorrect.
Furthermore, students can take high-quality data using a mobile phone camera, and subsequent analysis of the footage 
shows that the old simple description does not capture all of the details. To go beyond that description, one can use a computer algebra package to develop 
a more exact description of the system than the linearised approach given in the lab handout. This is done by explicitly including the equations of motion 
for the central mass that links the two pendula.  

We treat this mass as the bob of a third pendulum of mass $m_2$ and length $b$, where $m_2 \ll m$ (the mass of the main bobs) 
and we further treat all masses as point masses (We encourage readers to include a correction for the extended masses.). By using a Lagrangian 
approach, we  find equations of motion for all three masses in the system. This exposition is useful for students  studying Lagrangian dynamics and especially when looking for examples dealing with constrained systems.  It also demonstrates the advantages of using computer algebra packages in physics education as
many of the resulting formulae are too complicated to expect students to derive by hand.\cite{Savelsbergh.2000} 
\begin{figure}[h]
\vspace{0.5cm}
\centering
\includegraphics[width=7.5cm]{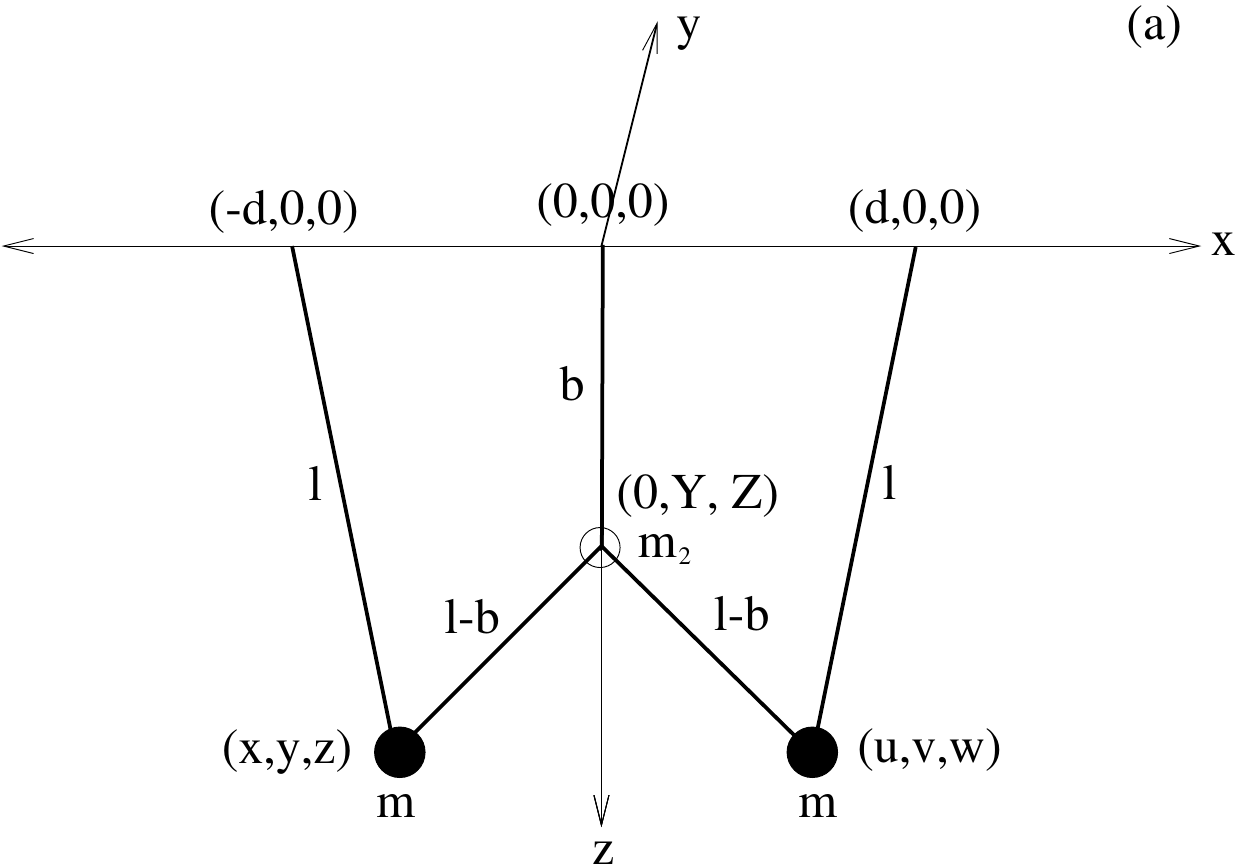}\hspace{1cm}
\includegraphics[width=5.0cm]{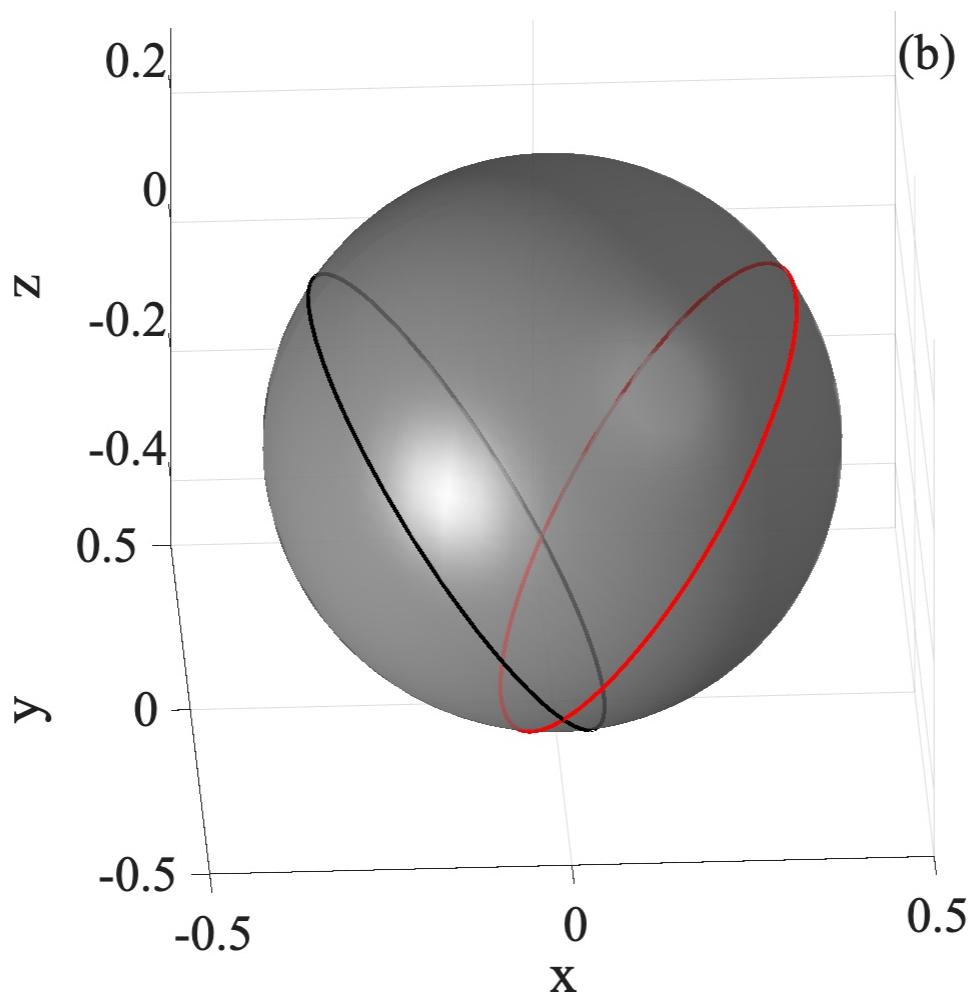}
\caption{(a) Schematic of the double pendulum with uneven supports. (b) (Color online) View of the possible positions in space for the pendulum 
bobs when the central pivot is at
$(0,0,-b)$. The red circle shows possible positions for the left bob while the black circle shows the possible positions for the right bob. 
Experimentally, the positions are limited to
lie near the bottom of the circle, which ensures that the small angle approximation remains valid.}
\label{schematic}
\end{figure}

To start the analysis, we use the co-ordinates and notation shown in Fig.~\ref{schematic}(a). There are eight variables of interest, namely $x(t),~y(t)$ and $z(t)$ for the 
position of the left pendulum bob, $u(t),~v(t)$ and $w(t)$ for the right bob, and $Y(t), Z(t)$ for the central  mass $m_2$. Note that $v(t)$ refers to the $y$-component of 
motion of the second bob, not the velocity of the first one. When stationary, each bob is a distance $l_z(b)$ from the horizontal $x$ axis where
\begin{equation}
l_z(b) = \frac{d^2 \sqrt{4 l (l-b)-d^2}+b \left(2 b l+d^2\right)}{2 \left(b^2+d^2\right)}.
\label{lz}
\end{equation}
Note that $l_z(0)=l_{\textrm{eff}}$ as defined earlier. 
In addition to the eight variables that describe the system, there are five constraints reflecting the fact that all of the strings have fixed lengths. These are:
\begin{align}
	\rm{C_1}: &\quad (x+d)^2+y^2+z^2 =l^2  \label{c1} \\
	\rm{C_2}: &\quad x^2+(y-Y)^2+(z-Z)^2 =(l-b)^2  \label{c2} \\
	\rm{C_3}: &\quad Y^2+Z^2 =b^2  \label{c3} \\
	\rm{C_4}: &\quad u^2+(v-Y)^2+(w-Z)^2 =(l-b)^2  \label{c4} \\
	\rm{C_5}: &\quad (u-d)^2+v^2+w^2 =l^2 . \label{c5} 
\end{align}
Each constraint reduces the number of degrees of freedom by one, so there are only three degrees of freedom in the problem. The impact of these can be 
understood by  first noting that each constraint describes a sphere in $\mathbb{R}^3$ and that the intersection of the two spheres is generically a circle. 
In particular, the  constraint $C_3$ describes a circle of radius $b$ centred at the origin, thus one degree of freedom ($\beta$) describes the position of the 
 central bob on this circle, i.e., $Y(\beta)  = b \sin(\beta) $ and 
 $     Z(\beta)  = -b \cos(\beta) $. Constraint $C_2$ describes a sphere of radius $l-b$ centred on the middle bob. On this sphere we first introduce the usual spherical
 co-ordinates except with the north pole pointing towards the pivot at $x=-d$. The constraint $C_1$ then forces the left pendulum bob to lie on a circle of constant
 latitude and so we introduce a second generalised co-ordinate $\phi$ to describe its position around this circle. Lastly, we re-parameterise the sphere so that the
 north pole points towards the pivot at $x=d$ and the constraint $C_5$ forces the right-hand bob to lie on a second circle of constant latitude described by a third
 generalised co-ordinate $\theta$. These two circles of constant latitude are shown in Fig.~\ref{schematic}(b) when the central bob is at the bottom of its swing.  
 As the central bob swings back and forth, the sphere described by constraint $C_2$ moves with it as do the circles of constant latitude on which lie the positions
 of the other two bobs. 

The cartesian position of the left bob written in terms of the generalised coordinates
is thus given by:
\begin{subequations}
\begin{alignat}{2}
x(\beta,\phi) &=&-\frac{2 b^2 d+b \cos (\phi ) \sqrt{-d^2 \left(4 l (b-l)+d^2\right)}-2 b d l+d^3}{2 \left(b^2+d^2\right)} \\
     y(\beta,\phi) & =& \quad \frac{ \sin (\beta ) \left(d \cos (\phi ) \sqrt{-d^2 \left(4 l (b-l)+d^2\right)}+b \left(2 b l+d^2\right)\right)}{2 \left(b^2+d^2\right)}  \nonumber \\
      &\, & -\frac{\cos (\beta ) \sin (\phi ) \sqrt{-d^2 \left(b^2+d^2\right) \left(4 l (b-l)\right)}}{2 \left(b^2+d^2\right)}  \\
    z(\beta,\phi) & = & -\frac{\cos (\beta ) \left(d \cos (\phi ) \sqrt{-d^2 \left(4 l(b-l)+d^2\right)}+b \left(2 b l+d^2\right)\right)}{2 \left(b^2+d^2\right)}  \nonumber \\
     &\, &      +   \frac{\sin (\beta ) \sin (\phi ) \sqrt{-d^2 \left(b^2+d^2\right) \left(4 l (b-l)+d^2\right)}}{ 2 \left(b^2+d^2\right)}.
        \end{alignat}
             \label{left}
\end{subequations}
A similar expression  applies for the right bob  with $\phi \rightarrow -\theta$. In practice, not all positions described by Eq.~\ref{left} are reachable 
due to the finite size of the bobs and physical nature of the supports. During the experiment, the students focus on measuring the frequencies 
of the normal modes, which means that small perturbations from equilibrium are considered and so collisions do not occur.

These generalized coordinates are then used to write the Lagrangian, which is
\begin{dmath}
L =1/2 \left[2 b^2 d^4 m - 4 d^6 m + 8 b^3 d^2 l m - 16 b d^4 l m +     16 b^4 l^2 m +  
8 b^2 d^2 l^2 m + 16 d^4 l^2 m + 8 b^6 m_2 +  
 16 b^4 d^2 m_2 + 8 b^2 d^4 m_2 +  
 4 b d^2 m \left(2 b l+d^2\right) \sqrt{4 l (l-b)-d^2} \left( \cos \theta +\cos \phi \right) +
 b^2 d^2 m \left(d^2- 4 l (l-b) \right)  \left( \cos \theta +\cos \phi \right) 
  \right] \dot\beta^2 + 
 2  m d\sqrt{- (b^2 + d^2) (d^2 + 4 l(b - l) )}    \left( d^2 \sqrt{-d (d^2 + 4 l(b - l))} +     b (d^2 + 2 b l) \right) 
 \left[  \cos\theta  \dot\theta -  \cos\phi   \dot\phi \right]  \dot\beta  +  
(b^2 + d^2) \left[4 g \left( \left(2 b (d^2 (m + m_2) + b (2 l m + b m_2)) + \\
m d^2  \sqrt{- (d^2 + 4 l(b - l) )}  (\cos\theta +   \cos\phi )\right) \cos\beta +  \\
m d  \sqrt{-(b^2 + d^2) (d^2 + 4 l(b - l) )}   (\sin\phi - \sin\theta) \right) \sin\beta  -  
m d^2 (d^2 + 4 l(b - l) )  (\dot\theta^2 +  \dot\phi^2 ) \right].
  \label{lagrangianfull}
   \end{dmath}

Solving the Euler-Lagrange equations gives 3 second-order nonlinear coupled equations for the evolution of $\phi, \, \theta$ and $\beta$. While this is conceptually straightforward, it requires further use of a computer algebra package due to the large numbers of terms involved. A second advantage of using a computer algebra package is that the equations of motion can also be numerically integrated, allowing  students to see if their resulting equations are physically reasonable (see the supplementary  information for examples).  

\subsection{Normal Modes of the System}
In the laboratory, students measure the periods of the different modes and the beat frequency, so we need to find the normal modes and their corresponding 
frequencies to facilitate comparison between  theory and experiment. As there are three generalised co-ordinates, there will be three normal modes, which 
stands in contrast to the two modes expected from a simple coupled pendulum. The previous lab manual gives explicit predictions for the frequencies of the 
even ($\omega_e$) and odd ($\omega_o$) modes, 
\begin{equation}
	\omega_e = \sqrt{\frac{g}{l}} \quad {\rm and} \quad \omega_o  = \sqrt{\omega_e^2+2 k}.
	\label{wrong}
\end{equation}
In 
the supplementary material
we show that for small angles the odd mode has a frequency  
\begin{equation}
  \omega_o=  \sqrt{\frac{2 g }{\sqrt{-d^2+4 l (l-b)}}}\, .
  \label{omegaa}
\end{equation}
The frequencies of the two even modes are more complicated but can be found straightforwardly. We begin by rewriting the three second-order coupled differential equations as six first-order nonlinear coupled differential equations in the form
\begin{equation}
\frac{d\,}{d t} {\bf x} = f({\bf x}),
\label{6thorder}
\end{equation}
where ${\bf x} = (\phi,\dot\phi,\theta,\dot\theta,\beta,\dot\beta)$. As expected, ${\bf x}=0$ is a fixed point of Eq.~\eqref{6thorder},  so linearising the equations about the fixed point means that we can approximate Eq.~\eqref{6thorder} by
\begin{equation}
\frac{d\,}{d t} {\bf x} = J {\bf x},
\label{jacobian}
\end{equation}
where $J$ is the Jacobian of $f$ evaluated at ${\bf x}=0$. The normal modes and their frequencies correspond to the eigenvectors and eigenvalues of $J$. 
The eigenvalues are the solutions of the characteristic equation ${\rm det}(J- \lambda I)=0$, which reduces to a cubic polynomial in $\lambda^2\equiv X$. 
Factoring out the term for the odd mode [Eq.~\eqref{omegaa}] results in a  quadratic equation in $X$ (given in the supplementary material), the roots of which 
can be found analytically. As expected, all three roots 
are negative, so  the frequencies 
of the modes are real and given by  $\omega_i = \sqrt{-X_i}$ where $X_i$ is the ith eigenvalue of the Jacobian. 
Again, it is not feasible for students to do this by hand.

As there are three normal modes, we denote the even mode with the highest frequency as the ``rapid'' mode while the other modes are called the slow even mode 
and the odd mode. Of the three modes, the odd mode is the easiest to explain: The middle bob is stationary and $\beta(t)=0$, implying that $Y(t)=0$. The other two bobs
move out of phase with each other at a frequency given by Eq.~\eqref{omegaa}. The slow even mode has $\phi(t), \theta(t) \ll \beta(t)$, and  all three bobs move in phase.
The last mode, the rapid even mode, has the central bob moving out of phase with the other two bobs and also with  greater amplitude. 
Experimentally,  the rapid even mode can be excited by displacing the mass $m_2$ by a small amount in the $+y$ direction; it can then be observed that the 
mass $m_2$ oscillates at a high frequency while the two much heavier masses are nearly stationary. In our experimental system the strings 
connecting the masses are made of fishing line and so have some elasticity. This means that the high frequency pendulum mode can easily be confused with a 
standing wave created in the string and so it is hard to say whether or not the high frequency mode is truly observed. We have not tried to replace the wire with 
either a heavier connector or a lighter one (such as a piece of thread) but we encourage readers to try this experiment. A fuller description of the modes is 
found in the supplemental material, where we also include Matlab code to numerically solve the full ODEs for arbitrary initial conditions.

\begin{figure}[h]
\vspace{0.5cm}
\centering
\includegraphics[width=15cm]{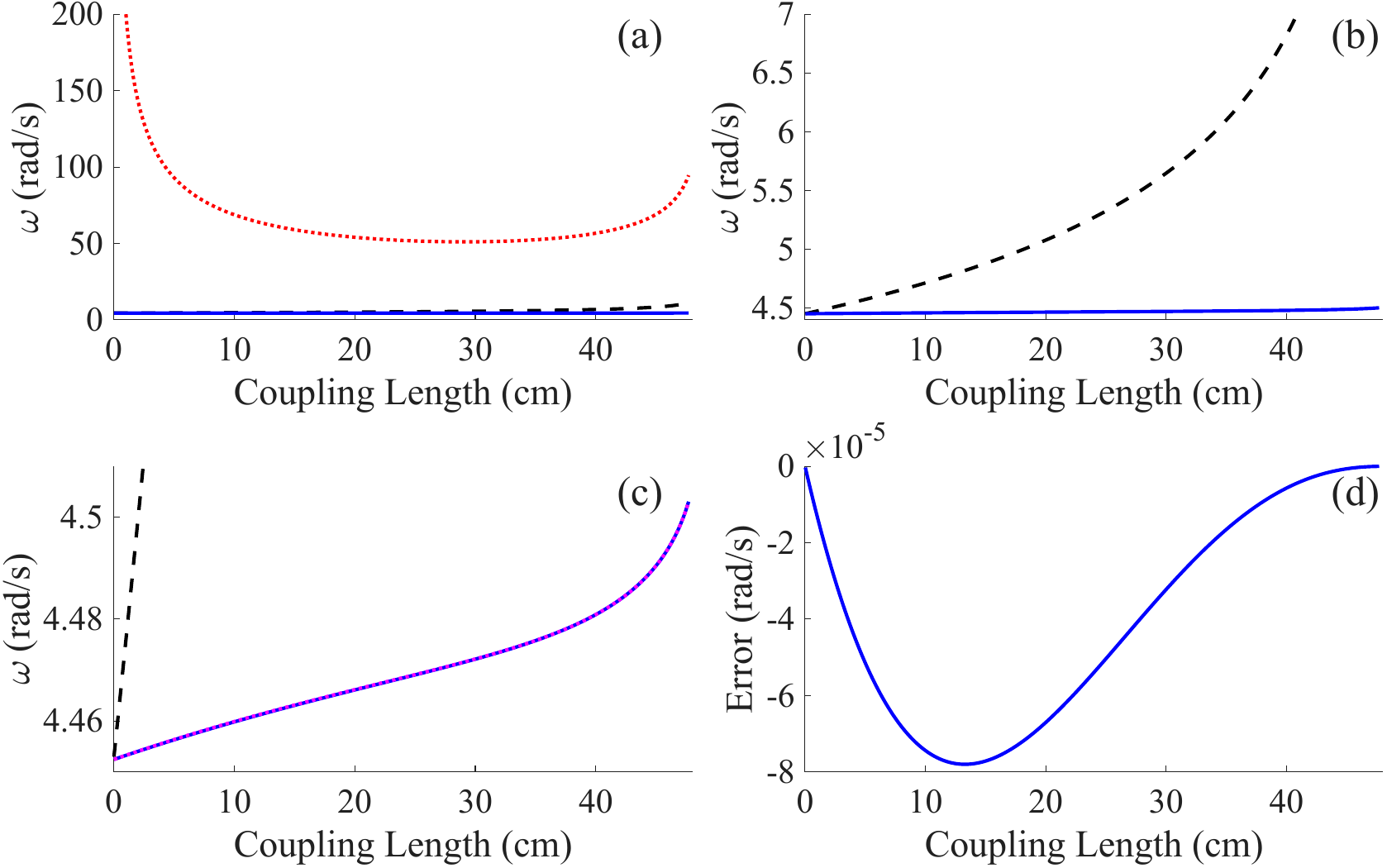} 
\caption{(Color online) (a) Frequencies of all of the modes of oscillations as a function of $b$  when $l=0.5\,\rm{m}$,  
$d=0.15\,\rm{m}$, $m=0.2\,{\rm kg}$, $m_2=0.005\,{\rm kg}$ and $g=9.8\, {\rm m\, s^{-2}}$ which are similar to those 
used in the lab. The rapid mode (red, dotted) has the highest frequency. The odd mode is the black (dashed) line. and the slow even mode is the blue
solid line. 
(b) A zoom in of (a) showing just the lowest two modes.
In (c) we zoom in further to focus on the slow mode and show $\omega_{e,\textrm{approx}}$ as a dotted line as well. On this scale the two are 
indistinguishable and lie on top of one another. Panel (d) shows the difference between this approximation for the slow even mode and the exact value. }
\label{freq}
\end{figure}

In Fig.~\ref{freq}(a) we show the frequencies of all three normal modes for the entire range of $b$. The rapid mode is shown as a  red dotted line, 
the odd mode as a black
dashed line, and the slow even mode as a solid blue line. The frequency of the rapid mode goes to infinity at $b=0$ and as  $b \rightarrow l$ 
since in these limiting cases the system 
reduces to either two identical  pendula or a single pendulum and thus the rapid mode does not exist. 
This frequency is always much greater than the other two; in order to show all three modes on the same scale we have set $m_2=5\,{\rm \, gm}$, which is greater than
the experimental value. The difference between the frequencies of the slow even mode and odd mode becomes apparent on this scale only for 
$b  \approx l$.  In Fig.~\ref{freq}(b) we zoom in and focus on the slow even mode and odd mode. For $b=0$ the two modes are identical, while 
for  $b\rightarrow l$ the frequency of the odd mode increases significantly while that of the slow even mode increases slightly.

In the third panel (bottom left), we zoom in again to focus on the slow even mode. Here it can be seen that, in contrast to the simple prediction in the lab manual, 
the frequency of  the slow even mode depends on the coupling length $b$, albeit to only a small degree. In addition to the exact solution we also show (red dotted curve) 
an approximate solution given by considering a compound pendulum of length $l_z$ with a mass of $2 m$ attached at the end and a mass $m_2$ attached at a 
distance of $b$ from the pivot. Such a pendulum has a moment of inertia of $m_2 b^2+ 2 m l_z^2$ about its pivot. For small angles the torque due to gravity is 
proportional to $m_2 g\, b+2 m g\, l_z$, so
its  oscillation frequency $\omega_{e,\textrm{approx}}$ is
\begin{equation}
	\omega_{e,\textrm{approx}} =\sqrt{\frac{m_2 g\, b+2 m g\, l_z(b)}{m_2 b^2+ 2 m [l_z(b)]^2}}.
	\label{freq_approx}
\end{equation}
Here  $l_z(b)$ is defined by Eq.~\eqref{lz} and  corresponds to the vertical displacement of the bobs from the $x$ axis  when they are stationary. 
The absolute error in the approximation is shown in Fig.~\ref{freq}(d). For all values of $b$ the error is less than one part in $10^{4}$, and goes to zero as
$b\rightarrow 0$ and $b\rightarrow l$ since in each case the system reduces to that of a simple pendulum. Note that the length of time  required  to see the 
difference between the improved  approximation, $\omega_{e,\textrm{approx}}$, and the actual frequency would be of the order of $10^4$ seconds and so 
 would not be  observable in practice.

\begin{figure}[h]
\centering
\includegraphics[width=15cm]{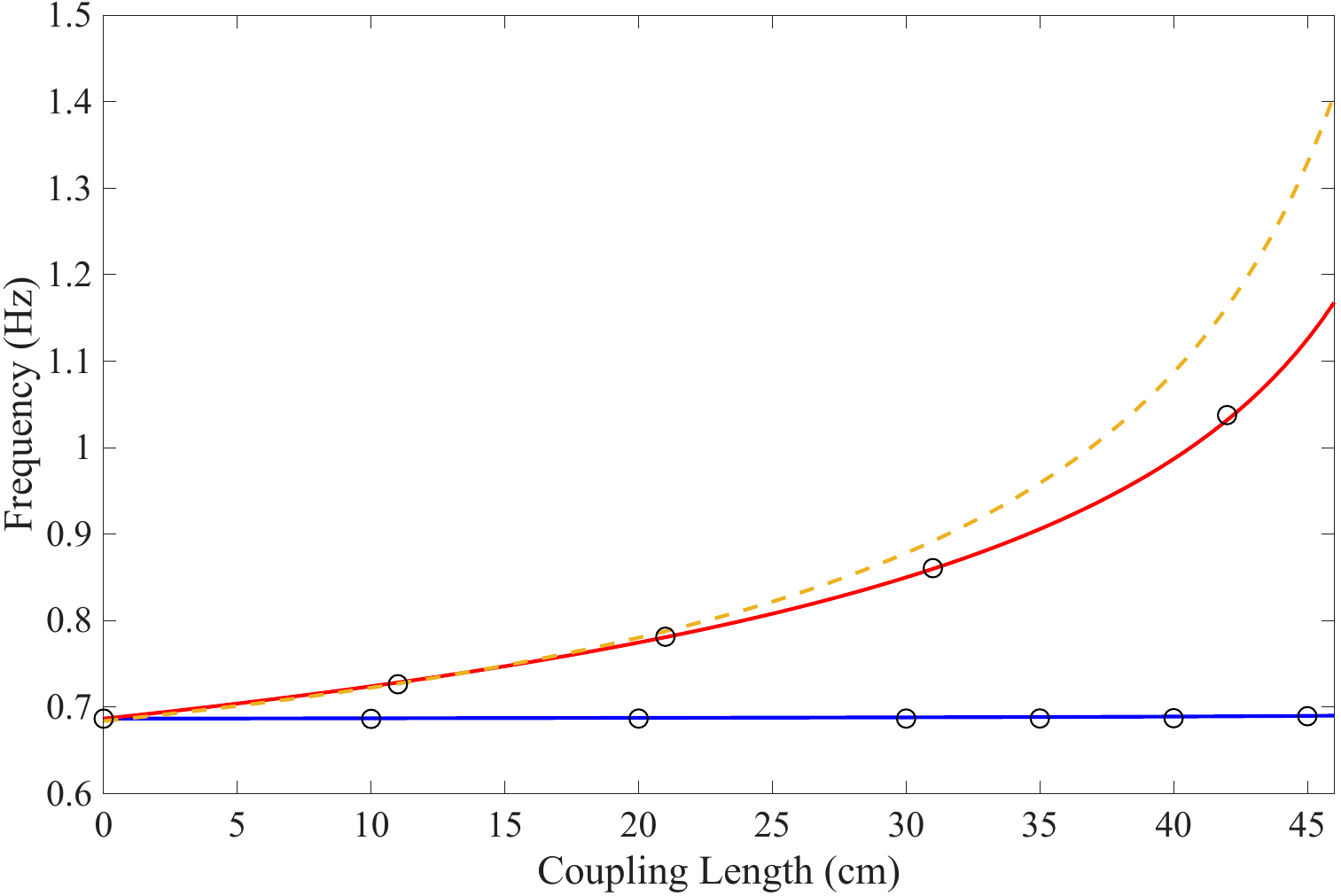} 
\caption{(Color online) The circles are the measured  frequencies of modes of oscillations as a function of $b$  with $l=0.531\,\rm{m}$,  
$d=0.135\,\rm{m}$, $m=0.2095\,{\rm kg}$, $m_2=0.001\,{\rm kg}$ and $g=9.8\, {\rm m\, s^{-2}}$. The odd mode is the top set of circles while the slow even
mode correspond to the lower set of circles. In addition we plot the improved theoretical
predictions for the modes (solid red and solid blue lines) and the previous prediction for the odd mode (dashed line) taken from Eq.~\ref{simple}. On this scale the 
slow even
mode appears constant but does in fact vary.}
\label{experimental}
\end{figure}

In Fig.~\ref{experimental} we show the measured frequencies as a function of the coupling length $b$ along with the theoretical predictions.
Here the measured values for the odd mode (which are the higher frequencies) fit well with our theoretical prediction (solid red line).
The old theoretical prediction given by Eq. 14 (shown as the dashed line) works reasonably well for small values of b but diverges at larger values.
 On this scale
the frequency of the slow even mode appears to be constant but in fact there is a small change that can be observed  using a video camera and data analysis software. 
Doing so, we found  a frequency of $0.6866\,{\rm Hz}$ for the uncoupled system and a period 
of $0.6893\,{\rm Hz}$ for tightly coupled pendulums. While this difference is small, playing the two
videos simultaneously does make it clear that it is real and that the pendulums are $\approx \pi$ radians
out of phase after 3 minutes. We note that the agreement between our predictions and the experimental result is perhaps surprising as we have not taken
into account the finite length of the pendulum bobs in the analysis, instead treating them as point masses. 
For a cylindrical mass of height $h$, the centre of mass 
lies at a distance of $h/2$ below its top, and so for our system this would increase the length of the pendulums by 1\,cm, reducing the frequency of 
oscillations by about  $1\%$. This change in frequency is much smaller than the change in frequency of the odd mode as $b$ is changed. 
For the even mode this change is about the same size as the effect of changing $b$, however it is a constant offset and does not effect the 
prediction that the frequency of the slow even
mode depends on the coupling length.  For a simple pendulum it has been shown that frequency change due to a finite sized mass is reduced
 when the mass 
of the string is also  included\cite{oliveira.2022}, and a similar effect might occur here explaining the agreement seen in Fig.~\ref{experimental}. 
This is something that could be studied in a future work to provide a more 
accurate model of  the system; alternatively, increasing the lengths of the strings and/or reducing the size of the masses would reduce the size of this error. 

\section{Extensions \label{extensions} }
Having discussed the experiment in its current state, it is worth briefly mentioning three possible extensions that can be used to stretch  students' understanding of the system and their technical capabilities. These extensions are all independent.

\begin{figure}[h]
\vspace{0.5cm}
\centering
\includegraphics[width=7.5cm]{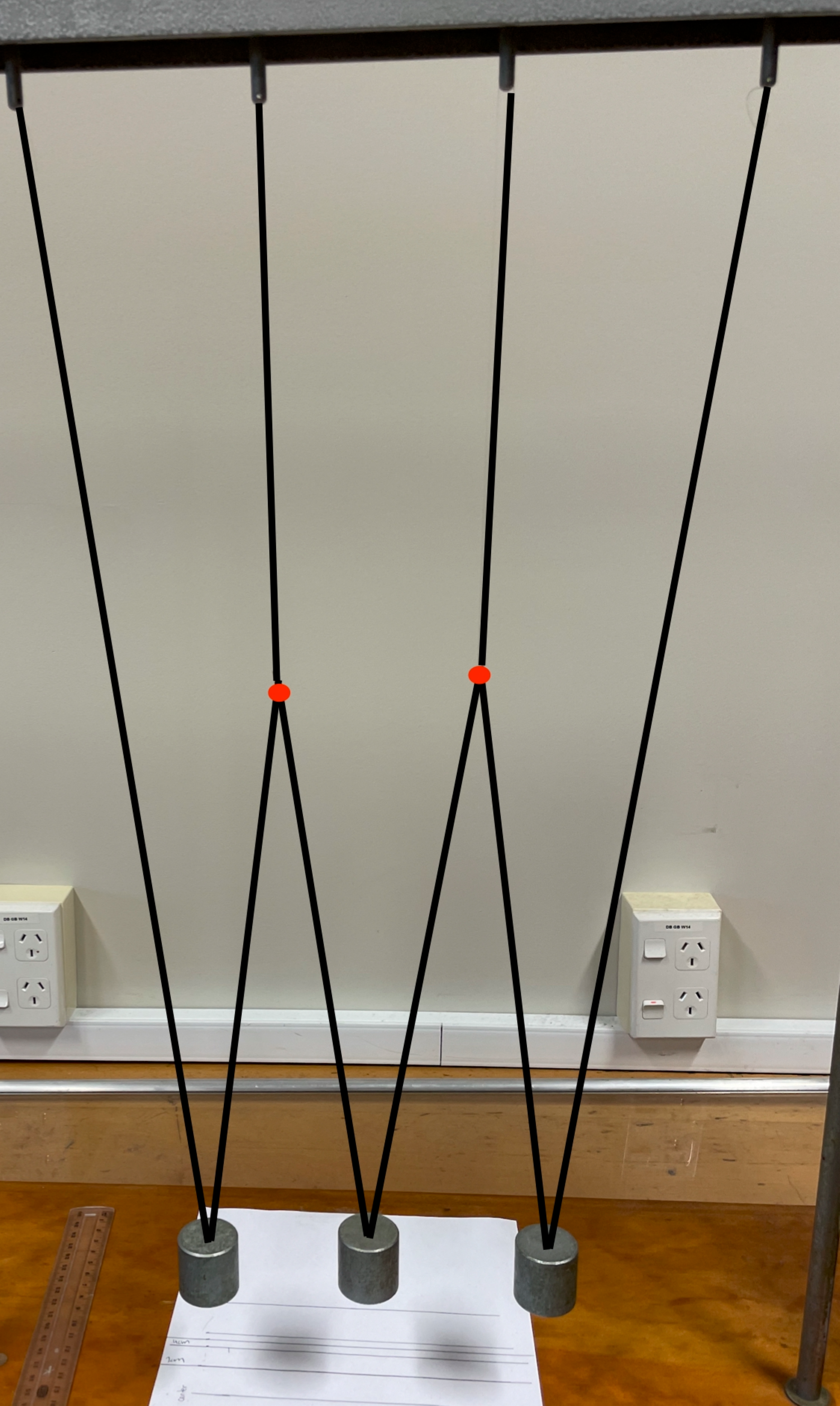} 
\caption{ (Color online) Picture of the triple pendulum. Superimposed on this we have added black lines over the
fishing lines and red dots over the two coupling points so that they are visible against the white
background.}
\label{figures/three}
\end{figure}

The first extension is to study more complicated systems. We currently ask students to predict the normal modes of the set of three coupled pendula as shown in 
Fig.~\ref{figures/three}. They are then asked to excite these modes and to observe them to see if they match their predictions. Our students are all able to find the 
normal modes  and can show that the relative amplitudes of the bobs match those of their predictions. This is a useful stepping stone to thinking about normal 
modes of large systems and how they can be used to simplify problems. This also highlights the flexibility of our approach  since adding more pendulums can be done 
far more cheaply and easily than with a system using, for example, coupled springs.

The second extension aims to strengthen students' analytic and computational abilities by having them use a computer algebra package to construct the 
Hamiltonian given the Lagrangian in Eq.~\eqref{lagrangianfull}. 
To do this, they need to identify the generalised momenta corresponding to the generalised coordinates, invert the equations so that the velocities can be 
written in terms of the momenta, and then write the Hamiltonian in terms of the momenta and generalised coordinates. 
Having written the Hamiltonian, students can then write the 
equations of motion for the pendulum bobs and solve them numerically to show that their results are correct by comparing with either their experiments or the 
results from the Euler-Lagrange equations.  The system is a six-dimensional Hamiltonian system and as such can be expected to behave chaotically in 
some regions and the students can explore this numerically as a computational task.

\section{Discussion and Conclusion}
We have presented a description of a coupled pendulum experiment currently being used in the undergraduate physics teaching labs at our University. 
Despite the apparatus having existed since at least the 1960s, its arrangement does not appear to be widely known. In comparison to other coupled 
pendulum experiments, ours has the advantage that the coupling strength can be easily varied, 
allowing students to directly observe the  effects on the normal modes. The apparatus itself comprises only two strings with weights and so can be easily 
assembled in virtually any lab from the high school level on up. 

Modeling this system analytically is out of reach for most  undergraduates, but computer algebra systems enable them to generate predictions from the full 
Lagrangian and to compare those predictions to their own data obtained through computer video analysis. This modern computer-aided approach allows students to 
engage with the relatively simple experimental apparatus at higher levels of understanding by giving them opportunities to experience experimental modeling and iteration 
to understand their results. Our experiment represents an inexpensive, accessible way for students to learn a range of important learning goals for beyond-first-year 
physics lab courses. 


\begin{acknowledgments}
The authors thank Bernd Krauskopf and Behrooz Yousefzadeh for helpful and illuminating discussions concerning this problem.
Generative AI was used to generate a preliminary draft of some sections of this article, based on outlines written by the authors.\cite{chatGPT} The authors corrected, edited, and significantly refined the AI draft output.
\end{acknowledgments}

\bibliographystyle{apsrev4-1}
\bibliography{pendulum_references}

\end{document}